\documentclass[prd,aps,eqsecnum,amsmath,floatfix,nofootinbib,preprint,tightenlines]{revtex4-2}

\usepackage{latexsym}
\usepackage{graphicx}
\usepackage[dvipsnames]{xcolor}

\usepackage{hyperref}

\def\bibi{\bibitem}

\def\b{\beta}
\def\c{\chi}

\def\f{\phi}                    
\def\g{\gamma}

\def\j{\psi}

\def\l{\lambda}
\def\m{\mu}
\def\n{\nu}

\def\p{\pi}                     
\def\th{\theta}                  
\def\s{\sigma}                  
\def\t{\tau}

\def\D{\Delta}

\def\L{\Lambda}

\def\U{\Upsilon}

\def\cl{{\cal L}}

\def\cs{{\cal S}}

\def\cbo{{\,\raise-.15ex\Sc [\,}}                       

\def\svev#1{\left\langle #1\right\rangle}       

\def\ddt#1{{\buildrel {\hbox{\LARGE .\kern-2pt.}} \over {#1}}}

\def\ie{\mbox{\it i.e.}}

\def\leqx{\,\raisebox{-1.0ex}{$\stackrel{\textstyle <}{\sim}$}\,}

\def\tr{{\rm tr}\,}

\def\half{{1\over 2}}

\def\ttl#1{{\it #1}}

\def\Osill{{O_{\rm sill}}}
\def\U{{\rm U}}
\def\SU{{\rm SU}}
\def\bj{\overline\j}

\def\hB{\hat{B}}
\def\tb{\tilde{b}}
\def\vevc{v}

\def\dChPT{\lowercase{d}C\lowercase{h}PT}

\begin{document}

\begin{center}
{\large\bf Power counting of the pion-dilaton effective field theory}\\[8mm]
Maarten Golterman$^a$ and Yigal Shamir$^b$\\[8 mm]
$^a$Department of Physics and Astronomy, San Francisco State University,\\
San Francisco, CA 94132, USA\\
$^b$Raymond and Beverly Sackler School of Physics and Astronomy,\\
Tel~Aviv University, 69978, Tel~Aviv, Israel\\[8mm]
\end{center}

\begin{quotation}
Confining QCD-like theories close to the conformal window
have a ``walking'' coupling. This is believed to lead to
a light singlet scalar meson in the low-energy spectrum, a dilaton,
which is the pseudo Nambu--Goldstone boson for the approximate scale symmetry.
Extending chiral perturbation theory to include the dilaton
requires a new small parameter to control the dilaton mass and its interactions.
In our previous work we derived a systematic power counting
for the dilaton couplings
by matching the effective low-energy theory to the
underlying theory using mild assumptions.  In this paper we examine two
alternative power countings which were proposed in the literature
based on a phenomenological picture for the conformal transition.
We find that one of these power countings fails, in fact, to
generate a systematic expansion; the other coincides with
the power counting we derived.
We also point out that the so-called $\D$-potential coincides
with the tree-level potential of the former, invalid, power counting.
\end{quotation}

\newpage
\section{\label{intro} Introduction}
QCD-like asymptotically free gauge theories in four dimensions
have two sources of explicit breaking of scale invariance.
The fermion mass provides a soft breaking, while the running
of the gauge coupling provides a hard breaking at the quantum level.

Asymptotic freedom restricts the number of Dirac fermions in the
fundamental representation, $N_f$, to be smaller than $\frac{11}{2}N_c$,
where $N_c$ is the number of colors.  It is generally believed that
this range is divided into two phases:
a confining and chirally broken phase\footnote{%
  The $N_f=1$ theory confines, but its only chiral symmetry $\U(1)_A$
  is anomalous.
}
for $2 \le N_f < N_f^*$,
and a phase with an infrared fixed point (IRFP), the conformal window,
for $N_f^* \le N_f \le \frac{11}{2}N_c$.
The critical value $N_f^*=N_f^*(N_c)$ is the sill of the conformal window.
The actual value of $N_f^*$ has been the subject of numerous lattice studies.
For the SU(3) gauge theory with $N_f=10$, a first nonperturbative calculation
of the beta function that covers the coupling range $g^2\,\leqx\, 20$
found an IRFP at $g_*^2 \approx 15$ \cite{Nf10}.
If confirmed by other studies, this would imply that $N_f^*(N_c=3) \le 10$.

When $N_f$ is close to, but below, the sill $N_f^*$, the beta function is small;
this is the region of a walking coupling.  When the theory
eventually confines, the smallness of the hard breaking of scale invariance
at the chiral-symmetry breaking scale is generally believed
to lead to the presence of a ``dilaton'' in the light hadron spectrum.
This is a flavor-singlet scalar meson,
which is then interpreted as a pseudo Nambu--Goldstone boson (NGB) arising from
the breaking of the approximate scale symmetry.
Notably, a light singlet scalar was found in lattice simulations
of the 8-flavor theory \cite{LSD1,LatKMI,LSD2,LSD3fit,LSD3num},
and of the sextet model
\cite{sextetconn,sextet1,sextet2,Kutietal,Kutietal20}.\footnote{
  It is unclear, however, if these theories are inside or below
  the conformal window.  The sextet model is an SU(3) gauge theory
  with two Dirac fermions in the sextet representations.  Strictly speaking,
  dChPT does not apply to this model since the Veneziano limit can be taken
  only for the fundamental representation.
}

In addition, the light meson spectrum contains the familiar pions,
which are the NGBs of spontaneously broken chiral symmetry.
Chiral perturbation theory (ChPT) is the well-founded effective field theory
(EFT) providing a systematic description of the pion sector.
The small parameter controlling the low-energy expansion is $m/\L$,
where $m$ is the fermion mass, and
$\L$ is the confinement scale of the (massless) theory.

In order to extend ordinary ChPT to includes the light dilatonic meson,
we need an additional small parameter to control the dilaton mass and its interactions.
What drives the smallness of the beta function
in the walking region is the proximity of the conformal window.
Hence, one expects the new small parameter to be proportional to
the distance to the conformal sill in theory space, $N_f^* - N_f$.
One can furthermore invoke the Veneziano limit \cite{VZlimit}
which turns this quantity
into a continuous parameter $n_f^*-n_f$, where $n_f=N_f/N_c$
and $n_f^* = \lim_{N_c\to\infty} N_f^*(N_c)/N_c$.

In a series of papers we developed \cite{GSEFT,latt16,gammay,LMP}
and applied \cite{GNS,KMIfit,NLO} dilaton chiral perturbation theory (dChPT),
an extension of ordinary ChPT which contains the dilatonic meson.
Any EFT should produce correlation functions that match those in the underlying
theory (order by order in the expansion in the small parameters of the EFT).
What this means is that concrete information about how scale invariance works
in the underlying theory is an essential ingredient of the EFT construction.
By matching the correlation functions of dChPT to the underlying theory
we showed that the new small parameter controlling the dilaton sector
is indeed $n_f^*-n_f$.  The power counting of dChPT is\footnote{%
  The power counting~(\ref{pcd}) applies in the Veneziano limit.
  Away from this limit one needs $1/N_c$ as an additional small parameter
  which controls the approach to the Veneziano limit.
  See also footnote~\ref{etaft} below.
}
\begin{equation}
\label{pcd}
p^2 \sim m \sim n_f^*-n_f \ .
\end{equation}

As we explain later on,
in comparison with ordinary ChPT, the theoretical foundations of dChPT
are less rigorous.  In particular, a few assumptions of a technical nature
are needed in order to set it up \cite{GSEFT,LMP}.  We have tested dChPT
by applying it to numerical data \cite{GNS,KMIfit,NLO}. However, it turns
out that such tests are, at present, limited by the available data.
It is thus important to continue scrutinizing the foundations of dChPT
also at the theoretical level.

The goal of this paper is to explore alternative power countings
that have been proposed in the literature as a starting point for the
construction of a low-energy EFT of pions and a dilatonic meson
and compare them with the power counting on which dChPT is based.
As we will discuss in more detail in the final section, a
different power counting will in general yield
a different EFT lagrangian at each order, and will thus lead to different
physical predictions. It is thus important to determine
the correct power counting for a pion-dilaton EFT.

One of the reasons why a pion-dilaton EFT is more complicated than
ordinary ChPT is the following.  Scale transformations act
on both fields and coordinates. As a result, a quartic potential for the
(effective) dilaton field is consistent with scale invariance.
Because the quartic potential does not violate any symmetry
of the pion-dilaton EFT, the power counting will necessarily allow
this potential to occur in the EFT with an $O(1)$ coupling.
That the quartic potential is not suppressed
by any small parameter is a fact that threatens the existence
of any systematic expansion for the pion-dilaton EFT,
and will turn out to play a key role.
Another important open question is what the physical mechanism is
that drives the conformal transition.  Ignorance about the correct answer
creates an additional difficulty when trying to identify
the correct power counting for the pion-dilaton EFT, as we will see.

This paper is organized as follows.  In Sec.~\ref{dchpt} we discuss dChPT.
We briefly review the derivation of its power counting~(\ref{pcd}),
with emphasis on
how the general principles for the construction of an EFT are applied.
We also explain how dChPT copes with the $O(1)$ quartic potential.

In Sec.~\ref{phenosill}, which contains the main results of this paper,
we begin by introducing a phenomenological picture for the onset
of the conformal transition \cite{GGS,CM}.
As it turns out, this physical picture
suggests {\em two} possible power countings (rather than a unique one)
for the pion-dilaton EFT, which we will refer to below
as the $G\to 0$ and the $\D\to 4$ power countings.
We show that  the $G\to 0$ power counting
actually fails to provide a systematic expansion.
The $\D\to 4$ power counting turns out to coincide with the power counting of dChPT.
Finally, we discuss the so-called $\D$-potential
\cite{LSD3fit,AIP1,AIP2,AIP3,AIPuniverse}, which
has been claimed \cite{AIP3,AIPuniverse,AIPLSD} to be
the leading-order dilaton potential arising from the $G\to 0$ power counting.
As a result of the failure of the $G\to 0$ power counting,
it follows immediately that the $\D$-potential fails to generate
a systematic expansion as well.  Once again,
the scale invariant quartic potential turns out to play a key role.
In Sec.~\ref{disc} we summarize and discuss our results.\footnote{%
  For other related low-energy approaches see Refs.~\cite{HLS,CT,RZ}.
}

As already mentioned, our approach is motivated by the presence
of a light singlet scalar in lattice simulations of walking theories.
The following caveat should, however, be borne in mind.
In the case of chiral symmetry, by applying the Goldstone theorem to the
{\em underlying theory} one proves that the pions are the NGBs arising
from the spontaneous breaking of chiral symmetry in the massless limit.
The low-energy theory must therefore contain the pions.
In the case of QCD with a small number of light flavors
there are no other parametrically light states, and
the low-energy theory contains the pions only; this leads to ordinary ChPT.
By contrast, it is unclear if the Goldstone theorem can be applied
to the underlying walking theories in the case of scale symmetry,
for several reasons.  First, reducing the hard breaking
of scale invariance requires ``motion'' in theory space.  This already involves
a double limit: the Veneziano limit followed by the limit where $n_f$
tends to $n_f^*$ from below.  In addition, the chiral limit becomes tricky
in a theory that is on the verge of developing an IRFP, as the fermion mass
becomes the only scale regulating the infrared upon entering
the conformal window.
Now, all the low-energy theories considered in this paper
are extensions of ordinary ChPT which have in common
an effective field that becomes the NGB of spontaneously broken scale symmetry
in the limit where scale invariance of the {\em low-energy theory}
becomes exact.  As it is not straightforward to establish a similar result
in the underlying theory, the presence of the dilaton field in
the effective low-energy theory has to be {\em postulated}.
In this paper, we will make this assumption, \ie, we will postulate that
walking theories produce
a light dilaton, governed by approximate scale symmetry.
As already mentioned above, the goal of this paper is then to study the
self-consistency of various power-counting schemes that have been proposed
for the low-energy pion-dilaton effective theory.

\section{\label{dchpt} \dChPT\ and its power counting}
Spurion fields play a pivotal role in the construction of an EFT.
The first step is to introduce the spurions into the underlying theory,
in such a way that the relevant symmetries are formally restored.
In the familiar case of ChPT one begins by promoting the
fermion mass $m$ to a matrix-valued spurion field $\c(x)$.
The fermion mass term
in the underlying theory then reads $\bj( P_R\,\c + P_L\,\c^\dagger ) \j$
where $P_{R,L}=\half(1\pm\g_5)$ are the chiral projectors.
With suitable transformation rules for the chiral spurion,
formal invariance under $\SU(N_f)_L\times\SU(N_f)_R$ is achieved.
Explicit breaking of the chiral flavor symmetry by the fermion mass is recovered by
setting the spurion field to its ``expectation value''
$\svev\c=\svev{\c^\dagger}=m$.\footnote{%
  We consider only the mass degenerate case to avoid irrelevant technicalities.
}
The full chiral symmetry is truly restored if we then take
the chiral limit $m\to 0$.

The EFT is constructed using effective fields for the low-energy
degrees of freedom, but in addition it must depend on the same set of spurions
that we have introduced into the underlying theory.
When we transform simultaneously the effective and the spurion fields,
the EFT is formally invariant under the same symmetries as the
underlying theory.  Explicit breaking of the relevant symmetries is recovered
once again by fixing the spurions to the same expectation values as
in the underlying theory.
All correlation functions that can be matched between the underlying theory
and the EFT are obtained by taking derivatives with respect to
the spurion fields.  Since the dependence of the underlying theory
on the spurions is always analytic by construction, so must be
the dependence of the EFT on the spurion fields as well.

Spurions can play a role in the power counting in two different ways.
In ordinary ChPT, the small expansion parameter is $m/\L$,
and the fermion mass $m$ is just the expectation value of the chiral spurion.
The underlying theory and the EFT both become chirally symmetric
by letting $m\to 0$.
But spurions can be involved in the power counting in a more subtle way as well.
As we discussed in Ref.~\cite{GSEFT}, the relevant example is the extension
of ordinary ChPT to include the $\eta'$ meson.

The $\eta'$ meson is the would-be NGB for the spontaneous breaking
of the axial $\U(1)_A$ symmetry.  Unlike the fermion mass, in any given theory
one cannot ``turn off'' the axial anomaly.  However, the axial anomaly
becomes parametrically small in the large-$N_c$ limit.\footnote{%
  Here the limit $N_c\to\infty$ is taken at fixed $N_f$.
  This is different from the Veneziano limit where the ratio
  $N_f/N_c$ is held fixed.
}
This leads to an extension of ChPT \cite{EW,KL}
that includes an effective field
for the $\eta'$ meson, which we will denote simply as $\eta(x)$.
At the level of the underlying theory,
the key step is to promote the vacuum angle $\th$ to a spurion field,
and to augment the gauge-field lagrangian by the term $i\th(x) q(x)$,
where $q(x)$ is the topological charge density
\begin{equation}
\label{theta}
q = \frac{c \l N_f}{N_c}\, \tr F\tilde{F} \ .
\end{equation}
Here $\l= g^2 N_c$ is the 't~Hooft coupling, and $c$ is a numerical constant.
The fact that $q(x)\propto 1/N_c$ (when expressed in terms of
the 't~Hooft coupling) means that it is parametrically small
in large-$N_c$ counting.

Under a $\U(1)_A$ transformation, both the effective field $\eta(x)$
and the spurion field $\th(x)$ transform by a common shift.
The shift is chosen such that the variation of the topological term
in the gauge-field lagrangian
cancels the anomalous $\U(1)_A$ transformation of the fermion determinant.
Unlike the case of the fermion mass, where chiral symmetry is restored
for $m\to 0$, here there is no fixed value for the vacuum angle $\th$
(including zero) that would lead to the restoration of $\U(1)_A$ invariance
in the underlying theory; the axial anomaly is always present.
Instead, differentiating the partition function of the underlying theory
with respect to $i\th(x)$ yields an insertion of the topological
charge density $q(x)$.  Since $q(x)$ is parametrically small for large $N_c$,
this ultimately allows for a systematic expansion for the EFT.\footnote{
  The power counting of the EFT is usually taken to be $p^2 \sim m \sim 1/N_c$.
}

The main difficulty that must be overcome at the level of the EFT is the
following. Because the $\eta$ and $\th$ fields are shifted by the same amount
under $\U(1)_A$ transformations, the difference $\eta-\th$ is invariant.
Hence, on symmetry grounds alone, any term in the EFT lagrangian
can be multiplied by a function of $\eta-\th$.
If that function was completely arbitrary, the EFT would lose predictability.
What allows for a systematic expansion in the EFT is that,
as in the underlying theory, each differentiation with respect to $\th$
must produce a factor of the small parameter $1/N_c$.
It follows that every function $V(\eta-\th)$ that occurs in the EFT
must have the Taylor expansion
\begin{eqnarray}
\label{VNc}
V &=& \sum_{n=0}^\infty c_n (\eta-\th)^n \ ,
\\
c_n &=& O(1/N_c^n) \ .
\nonumber
\end{eqnarray}
At each order in the low-energy expansion of the EFT,
the hierarchy of the $c_n$ coefficients implies that we must truncate
the Taylor series to a finite number of terms.
In particular, at leading order, the functions of $\eta-\th$
that multiply the pion kinetic term and the pion mass term
in the tree-level lagrangian must be truncated to their leading
constant term $c_0$, and can thus be disregarded altogether.
Once all the relevant functions of $\eta-\th$ have been
truncated appropriately, one can set $\th=0$ and proceed with the calculation
of correlation functions to a given order in the low-energy expansion.

The systematic expansion of dChPT works in a similar way.\footnote{%
  For full details, see Ref.~\cite{GSEFT}.
}
Much like the $\eta$ and $\th$ fields associated with the $\U(1)_A$ symmetry,
we introduce a spurion field $\s(x)$ and an effective dilatonic meson
field $\t(x)$, both dimensionless,
that transform by a common shift under dilatations.  To be precise,
\begin{subequations}
\label{scale}
\begin{eqnarray}
\label{scales}
\s(x) &\to& \s(\l x) + \log\l \ ,
\\
\label{scalet}
\t(x) &\to& \t(\l x) + \log\l \ .
\end{eqnarray}
\end{subequations}
Using dimensional regularization
the bare action of the underlying theory takes the form
\begin{equation}
\label{Sbare}
S = \m_0^{d-4} \int d^dx\, e^{(d-4)\s(x)} \cl_0(x) \ .
\end{equation}
The bare lagrangian $\cl_0(x)$ is defined in terms of bare fields
and couplings that all transform according to their canonical
mass dimension.\footnote{
  To avoid irrelevant technicalities, we set the fermion mass to $m=0$
  throughout most of this section.
}
The mass dimension of $\cl_0(x)$ is 4,
and it thus transforms under dilatations as $\cl_0(x) \to \l^4\cl_0(\l x)$.
The fixed reference scale $\m_0$ does not transform under dilatations
by construction,\footnote{%
  See, however, the generalized discussion in Ref.~\cite{gammay}.
}
and can be thought of as an ultraviolet cutoff.
Thanks to the spurion transformation rule~(\ref{scales}),
the action~(\ref{Sbare}) is formally invariant under dilatations.

Much like the case of the axial anomaly, setting $\s(x)$ to any fixed value
will not restore scale invariance of the underlying theory;
it is always broken at the quantum level by the running of the coupling.
However, differentiating the underlying partition function with respect to
$\s(x)$ again produces an insertion of a parametrically small operator.
Expressing the action in terms of renormalized fields and couplings,
it takes the form\footnote{%
  Equation~(\ref{Sren}) is true up to local operators that are multiplied by
  higher powers of $\s(x)$, that do not affect the construction
  of the EFT \cite{GSEFT}.
}
\begin{equation}
\label{Sren}
S = \int d^dx\, \Big( \cl_{\rm ren} + \s T \big) \ ,
\end{equation}
where $\cl_{\rm ren}$ is the usual renormalized lagrangian, and
\begin{equation}
\label{Tan}
T = \frac{\b(g)}{2g}\, F^2
\end{equation}
is the trace anomaly \cite{CDJ}.
Here $\b(g) = \m (\partial g / \partial\m)$ is the beta function, and
$F^2 = F_{\m\n}^a F_{\m\n}^a$.  It follows that the differentiation
with respect to $\s(x)$ yields an insertion of the trace anomaly,
which is parametrically small for walking theories close
to the sill of the conformal window,
\begin{equation}
\label{Tnf}
T \sim n_f^*-n_f \ .
\end{equation}

In analogy with the case of $\U(1)_A$ transformations,
any function $V$ of the field difference $\t(x)-\s(x)$ transforms homogeneously
under dilatations, $V(x)\to V(\l x)$.
Thus, $V(\t-\s)$ can multiply any existing term in the EFT
which is already allowed by the symmetries.
As before, this could lead to the loss of predictability, but again,
the requirement that each differentiation with respect to $\s(x)$
generates one power of the small parameter $n_f^*-n_f$
implies that any such function occurring in the lagrangian of dChPT
must satisfy (compare Eq.~(\ref{VNc}))
\begin{eqnarray}
\label{Vnf}
V &=& \sum_{n=0}^\infty c_n (\t-\s)^n \ ,
\\
c_n &=& O((n_f^*-n_f)^n) \ .
\nonumber
\end{eqnarray}
Thus, the function $V(\t-\s)$ that multiplies
a particular term in the EFT must again be truncated at a finite order of its
Taylor series, according to the order of the low-energy expansion
we are interested in.  In particular, at leading order in the low-energy
expansion, which is always $O(p^2)$,
the functions that multiply the pion and dilaton kinetic terms
in the tree-level dChPT lagrangian (as well as the function that multiplies
the pion mass term when $m\ne 0$) must be truncated to their
leading constant term, and can be disregarded once again.

The only exception is the pure dilaton potential, to which we now turn.
First, the quartic dilaton potential discussed in the introduction takes
the form $e^{4\t}$.  It transforms as $e^{4\t(x)} \to \l^4 e^{4\t(\l x)}$,
and thus the corresponding term in the action is indeed scale invariant.
Next, according to the previous discussion, this term should be multiplied
by a function of $\t-\s$, which in turn should be truncated appropriately.
Finally we set $\s=0$.
The pure-dilaton part of the tree-level lagrangian of dChPT is
\begin{equation}
\label{Ld}
\frac{f_\t^2}{2}\, e^{2\t} (\partial_\m \t)^2 + V_d \ .
\end{equation}
We will shortly see how the parameter $f_\t$ is related to
the physical dilaton decay constant.  Guided by the power counting~(\ref{pcd}),
and remembering that the tree-level lagrangian of the EFT is always $O(p^2)$,
it follows from the discussion above that
the resulting tree-level dilaton potential can be expressed as
\begin{subequations}
\label{VUd}
\begin{eqnarray}
\label{Vd}
V_d &=& f_\t^2 B_\t U \ ,
\\
\label{Ud}
U &=& e^{4\t} (c_0 + c_1 \t) \ ,
\end{eqnarray}
\end{subequations}
where $c_0=O(1)$ and $c_1=O(n_f^*-n_f)$.  The parameter $B_\t$
has mass dimension 2.  Note that the pure $e^{4\t}$ potential
indeed occurs with an $O(1)$ coupling, in agreement with our general reasoning.

Solving the classical equation, the minimum of the potential is
\begin{equation}
\label{vd}
v = \svev\t = -\frac{1}{4} - \frac{c_0}{c_1} \ ,
\end{equation}
and the curvature at the minimum is
\begin{equation}
\label{curve}
U'' = 4 c_1 e^{4v} \ ,
\end{equation}
where $v$ is given by Eq.~(\ref{vd}).  Now, virtually all terms occurring
in the dChPT lagrangian involve some power of $e^\t$.
As a result, low-energy constants get ``dressed'' by a corresponding
power of $e^v$.  In particular, it follows from Eq.~(\ref{Ld}) that
the dilaton decay constant\footnote{%
  Since $m=0$ this is the dilaton decay constant in the chiral limit.}
is $F_\t = f_\t e^v$.
The dressed, physical $B_\t$ parameter is $\hB_\t = B_\t e^{2v}$.
The dilaton mass is given by
\begin{equation}
\label{Mdsq}
M_\t^2 = \frac{V_d''}{F_\t^2} = 4 c_1 \hB_\t \ .
\end{equation}

Equation~(\ref{Mdsq}) implies that the dilaton mass (squared)
is parametrically small, because $c_1$ is, as expected for a pseudo NGB.
This has the following immediate implication.  The dimensionless ratio
which controls the low-energy expansion in the dilaton sector is (for $m=0$)
\begin{equation}
\label{tauratio}
\frac{M_\t^2}{(4 \pi F_\t)^2} =  \frac{4 c_1 \hB_\t}{(4 \pi F_\t)^2}
=  \frac{4 c_1 B_\t}{(4 \pi f_\t)^2}\ .
\end{equation}
This ratio is proportional to $c_1$, making it parametrically small
as well, as it must be for a systematic low-energy expansion.
A similar behavior is found when the conformal sill
is approached at nonzero $m$.
For any given $m>0$, when $n_f \to n_f^*$ one enters a hyperscaling regime
where \cite{LMP}
\begin{equation}
\label{ratiohscale}
\frac{M_\t^2}{(4 \pi F_\t)^2} \;\sim\; c_1 \log(c_1)
\;\propto\; (n_f^*-n_f) \log(n_f^*-n_f) \ .
\end{equation}
Thus, the ratio controlling the low-energy expansion in the dilaton sector
tends to zero when the conformal sill is approached.

Returning to $m=0$,
of the four parameters that occur in the tree-level lagrangian
of the dilaton sector, the ratio~(\ref{tauratio}) depends on $f_\t$, $B_\t$
and $c_1$.  But it is independent of $c_0$, the troublesome $O(1)$ coefficient
of the quartic $e^{4\t}$ potential.  This remarkable property is intimately
related to a unique feature of dChPT: the freedom to shift
the $\t$ field by a constant.

As we have discussed extensively
in Refs.~\cite{GSEFT,LMP}, the effect of shifting the $\t$ field can be absorbed
into redefinitions of the parameters of the theory.
The behavior of the parameters of the pure dilaton sector
is as follows.  $f_\t$, $B_\t$ and $c_0$ are affected
by the $\t$-shift, while $c_1$ is invariant.  The dressed parameters
$F_\t$ and $\hB_\t$ are invariant as well.
Thus, the dilaton mass~(\ref{Mdsq}) and the ratio~(\ref{tauratio})
are invariant under the $\t$-shift, as all physical quantities must be.
On the other hand, $c_0$ is affected by the $\t$-shift,
and this explains why the dimensionless ratio~(\ref{tauratio})
cannot contain $c_0$.  In the next section we will see
that the situation is different for another power counting
that has been proposed for the pion-dilaton EFT.

The above discussion requires one refinement.  Because $c_0=O(1)$
while $c_1=O(n_f^*-n_f)$, the expectation value $v$ of the dilaton field
in Eq.~(\ref{vd}) is parametrically $O(1/(n_f^*-n_f))$.
Considering the Taylor series~(\ref{Vnf}) for the case of the dilaton potential,
if we substitute $\t=v$ (and $\s=0$) then all terms in the series
become $O(1)$ within the power counting~(\ref{pcd}).
As a result, certain mild technical assumptions about the global features
of the dilaton potential are required in order to set up dChPT.
The most important one is that, taken as a function of the product
$(n_f^*-n_f)\t$, the dilaton potential vanishes for some value of its argument
\cite{CLPRV}.
Once this condition is satisfied, there exists a $\t$-shift that
eliminates the $O(1)$ part of $c_0$.  After that, the new value of $c_0$ is
$O(n_f^*-n_f)$, at which point the previous discussion applies.\footnote{%
  An additional $\t$-shift can be done to set the $O(n_f^*-n_f)$ part of $c_0$
  to some desired value, see Ref.~\cite{LMP}.
}
This is proven in detail in Appendix~A of Ref.~\cite{LMP}.

In summary, having made the basic assumption that the
underlying walking theory produces a light dilaton,
and thus that the low-energy EFT
should contain a dilaton field, the power counting of dChPT is derived
by matching its correlation functions to the underlying theory.
Thanks to its behavior under $\t$-shifts,
at the price of the additional technical assumptions mentioned above
dChPT copes with the $O(1)$ coefficient of the quartic potential
and successfully generates a systematic low-energy expansion.

We conclude this section with a technical comment.  Non-perturbatively,
the underlying non-abelian theory is always defined on the lattice,
where the breaking of spacetime symmetries,
including in particular dilatations, takes a complicated form.
However, the derivation of an EFT always involves the intermediate step
of first constructing the Symanzik effective theory, which lives
in the continuum.  Thus, the bare continuum theory~(\ref{Sbare}) can be
identified with the (dimensionally regularized) Symanzik effective theory
at leading order in an expansion in the lattice spacing.

\section{\label{phenosill} Phenomenology of the conformal transition\\
and related power countings}
In this section we turn to alternative power countings for the
pion-dilaton EFT which have been proposed in the literature.
As explained in the introduction, in all cases it is assumed
that, besides pions, the low-energy EFT contains a dilaton field:
an effective field that becomes the NGB of spontaneously broken scale invariance
in the limit where scale invariance of the underlying walking theory becomes exact.

The starting point is a certain physical picture for the mechanism that drives
the transition into the conformal window.
According to this physical picture, (massless) theories near the sill of the
conformal window are governed in the infrared
by a phenomenological lagrangian of the form  \cite{GGS,CM}
\begin{equation}
\label{totL}
\cl = \cl_{\rm conf} + G\Osill \ .
\end{equation}
By assumption, $\cl_{\rm conf}$ describes a conformal theory.  Hence, in itself
it would be suitable to describe the low-energy physics of
a theory inside the conformal window,
whose infrared behavior is governed by an IRFP.  The conformal transition
is then attributed to the presence of $\Osill$ in the lagrangian.
Under a scale transformation, we assume that
\begin{equation}
\label{Oscl}
\Osill(x) \to \l^\D\, \Osill(\l x) \ .
\end{equation}
Since the lagrangian~(\ref{totL}) is in itself an effective description
of the underlying theory, the coupling $G$ and the scaling dimension $\D$
of $\Osill$ are not free parameters.  Rather, they are both
functions of $N_c$ and $N_f$.

This starting point raises numerous questions.  A crucial question
is whether the phenomenological lagrangian~(\ref{totL}) can be derived from
the standard renormalizable lagrangian that defines the underlying gauge theory.
We will defer all such questions to the discussion section,
whereas in this section we will discuss the implications
of this physical picture for the power counting of the pion-dilaton EFT.

As was realized in Refs.~\cite{GGS,CM}, there are two ways by which conformality of
the phenomenological lagrangian in Eq.~(\ref{totL}) can be restored,
and this suggests two alternative power countings for the pion-dilaton EFT:

\smallskip\noindent
{\em $G\to 0$ power counting.}  Here one envisages that $G>0$ below the
conformal sill while $G=0$ at (and above) the sill,
where the theory has an IRFP.
By assumption, $G$ tends to zero as the sill is approached from below
in flavor space.
The scaling dimension $\D$ is kept as a free parameter.
The corresponding power counting is
\begin{equation}
\label{pcG}
p^2 \sim m \sim G\ .
\end{equation}
We will discuss this power counting in Sec.~\ref{Gto0}.
We note that, like the fermion mass $m$, also the coupling $G$ is
dimensionful.  Hence, the dimensionless expansion parameters
are $m/\L$ and $G/\L^{4-\D}$.  The infrared scale $\L$ may be
identified with the vacuum expectation value (VEV) of the effective dilaton field $\c$
that will be introduced in Sec.~\ref{Gto0} below.
The main result of this paper is that the $G\to 0$ power counting
actually fails to provide a systematic expansion.

\smallskip\noindent
{\em $\D\to 4$ power counting.}  Alternatively,
one assumes that $G$ is $O(1)$ on both sides of the conformal sill.
Instead, it is assumed that the scaling dimension $\D$ tends to 4 at the sill.
Usually it is further assumed that $\D<4$ below the sill, so that  $\Osill$ is
a relevant operator, while $\D>4$ above the sill, and thus $\Osill$ is
irrelevant at the IRFP.  At the sill $\D=4$, and $\Osill$
is marginal (at the quantum level), thereby restoring scale invariance
when the conformal sill is approached from below.
The power counting is then
\begin{equation}
\label{pcD}
p^2 \sim m \sim |4-\D| \ .
\end{equation}
We discuss this option in Sec.~\ref{Dto4}.
Finally, in Sec.~\ref{smr} we offer a brief summary,
while in Sec.~\ref{Dpot} we discuss the so-called $\D$-potential
\cite{LSD3fit,AIP1,AIP2,AIP3,AIPuniverse,AIPLSD}.

According to an interesting scenario \cite{KLSS,GRZ}, what sets the sill
of the conformal window is a combination of one UV and one IR real fixed point
which exists inside the conformal window, which then ``collide'' at the sill,
and finally move into the complex plane in the ``walking'' regime.\footnote{
  For an alternative scenario, see Ref.~\cite{latt16}.}
It has been suggested that the colliding fixed-points scenario
does not necessarily require the existence of a parametrically light dilaton
in walking theories \cite{GRZ}.  We note, however, that
the colliding fixed points scenario is perfectly compatible with
a low-energy pion-dilaton EFT endowed with the $\D\to 4$
power counting, as it
envisages the presence of an operator whose scaling dimension $\D$ behaves
in the same way as assumed for the $\D\to 4$ power counting.
By contrast, the scenario is not compatible with the $G\to 0$ power counting,
because this power counting treats $\D$ as a free parameter, whereas
the colliding fixed points scenario requires the existence of an
operator that becomes exactly marginal at the conformal sill.
Moreover, as we will show in Sec.~\ref{Dto4}, the $\D\to 4$ power counting
leads us naturally to dChPT.  Thus, if walking theories have
a light dilaton, as suggested by numerical simulations,
it follows that dChPT is the correct EFT
for the colliding fixed point scenario as well.

\begin{boldmath}
\subsection{\label{Gto0} $G\to 0$ power counting}
\end{boldmath}
Using $G$ as a small parameter works in a way similar to the familiar
chiral expansion.  Taking the phenomenological lagrangian~(\ref{totL})
to represent the underlying theory, one begins by promoting the coupling $G$
to a spurion field $\cs(x)$.  The lagrangian becomes\footnote{%
An implicit assumption is that $\cl_{\rm conf}$ is perturbed
by a {\em single} operator $\Osill$.  If several operators $O_{\rm sill}^i$,
$i=1,2,\ldots,$ are required for the phenomenological lagrangian,
each operator would be accompanied by a separate spurion,
which in turn would ultimately alter the low-energy EFT.}
\begin{equation}
\label{totLS}
\cl = \cl_{\rm conf} + \cs\Osill \ ,
\end{equation}
and with the spurion transformation rule
\begin{equation}
\label{Sstrans}
\cs(x) \to \l^{4-\D}\,\cs(\l x) \ ,
\end{equation}
the lagrangian is formally invariant under dilatations.
The original lagrangian, in which scale invariance is explicitly broken
by $\Osill$, is recovered by fixing the spurion field to $\svev\cs=G$.

As in Sec.~\ref{dchpt}, at the level of the EFT we will focus
on the pure dilaton sector, setting $m=0$.  Its lagrangian is
\begin{equation}
\label{pured}
\cl_d = \half \partial_\m\chi \partial_\m\chi + V(\c) \ .
\end{equation}
The scale transformation of the dilaton field $\c(x)$ is
\begin{equation}
\label{transchi}
\c(x) \to \l\, \c(\l x) \ .
\end{equation}
The dilaton potential has the expansion
\begin{equation}
\label{Vbn}
V = \c^4 \sum_{n=0}^\infty b_n \Big(\cs \c^{\D-4}\Big)^n \ .
\end{equation}
The low-energy constants $b_n$ are invariant under dilatations,
and they are $O(1)$ in the power counting, just like the
low-energy constants in standard ChPT.
With the transformation rules of the dilaton and spurion fields
introduced above, the EFT action defined by the
lagrangian~(\ref{pured}) is formally invariant under dilatations.

After fixing the spurion to $\svev\cs=G$ the potential becomes
\begin{equation}
\label{VbnG}
V = \c^4 \sum_{n=0}^\infty b_n \Big(G\c^{\D-4}\Big)^n \ ,
\end{equation}
making the expansion in $G$ manifest.  At the same time,
since the spurion has been set to a fixed value,
the potential breaks scale invariance explicitly.
As usual, the exception is the leading term in the sum:
this is the by-now familiar quartic potential $\c^4$,
whose coupling $b_0$ is parametrically $O(1)$.

According to the power counting~(\ref{pcG}),
at tree level we keep terms up to $O(G)$, hence
\begin{equation}
\label{Vdn}
V = b_0 \c^4 - \tb_1 \c^\D \ ,
\end{equation}
where we have introduced $\tb_1=-b_1 G$ (the minus sign is for convenience).
We will assume that both $b_0$ and $\tb_1$ are positive.
We also require $0<\D < 4$. The upper bound $\D < 4$ is needed to ensure
boundedness from below of the potential.  The lower bound $0<\D$
is to avoid a singularity of the potential for $\c\to 0$.\footnote{
Additional assumptions are needed for general, non-integer, $\D$,
where the potential is not well-defined for negative $\c$.
Technical assumptions of a similar nature that are needed in dChPT
are discussed in Appendix~A of Ref.~\cite{LMP}.}

We comment that, in analogy with the cases discussed in Sec.~\ref{dchpt},
every term in the lagrangian of the EFT which is
(a) allowed by the symmetries, and (b) independent of the $\cs$ spurion,
should be multiplied by a function $U=U(\cs \c^{\D-4})$.
Actually, the dilaton potential~(\ref{Vbn}) has this form, where the
original potential allowed by the symmetries is the pure $\c^4$.  Similarly,
the dilaton kinetic term in Eq.~(\ref{pured}) should be multiplied by
another function of the argument $\cs \c^{\D-4}$,
which, after setting $\svev\cs=G$,
becomes a function of the argument $G \c^{\D-4}$.  At tree level, however,
that function would again have to be truncated to its constant leading term.
Hence, we omitted it from Eq.~(\ref{pured}).

We expand the dilaton field as
\begin{equation}
\label{chiexpand}
\c = v + \f,
\end{equation}
where the VEV is $\svev\c=v$, and $\f$ is the quantum part.
The saddle-point equation can be written as
\begin{equation}
\label{chivevv}
0 = v V'(v) = 4b_0 v^4 - \D\,\tb_1 v^\D \ .
\end{equation}
The absolute minimum is the $v>0$ solution,
\begin{equation}
\label{vsolve}
v^{4-\D} = \frac{\D}{4}\frac{\tb_1}{b_0} \ .
\end{equation}
Upon substituting the field expansion~(\ref{chiexpand})
into the potential~(\ref{Vdn}) and using the classical solution~(\ref{vsolve}),
the potential becomes
\begin{equation}
\label{Vexpand}
V = \frac{\D-4}{\D}\, b_0 v^4 + \frac{M_\t^2}{2}\, \f^2
    + M_\t^2 v^2 \sum_{n\ge 3} g_n \frac{\f^n}{v^n} \ ,
\end{equation}
where the dilaton mass squared is\footnote{%
Equations~(\ref{vsolve}) and~(\ref{Md}) were previously derived in Ref.~\cite{CM}.
}
\begin{equation}
\label{Md}
M_\t^2 = 4 b_0 v^2 (4-\D) \ .
\end{equation}
The first few coefficients in the sum in Eq.~(\ref{Vexpand}) are
\begin{eqnarray}
\label{gn}
g_3 &=& (\D+1)/3! \ ,
\\
g_4 &=& (\D^2-2\D+3)/4! \ ,
\nonumber\\
g_5 &=& (\D-1)(\D-2)(\D-3)/5! \ ,
\nonumber\\
g_6 &=& (\D-1)(\D-2)(\D-3)(\D-5)/6! \ ,
\nonumber
\end{eqnarray}
and so on.  Note that the terms with $n\ge 5$ arise from the
Taylor series of $\c^\D=(v+\f)^\D$ only, hence all these terms
contain an explicit factor of $\D-4$.  After we absorb this factor into $M_\t^2$
it follows that $g_n$ tends to a non-zero value in the limit $\D\to 4$.

The reader who is familiar with ordinary ChPT
will recognize that the expansion in Eq.~(\ref{Vexpand}) is
qualitatively similar to the expansion of the pion mass term in
the standard chiral lagrangian, if the identifications $\f\leftrightarrow\p$
and $v\leftrightarrow F_\p$ are made, where $\p$ is the pion field
and $F_\p$ its decay constant.  In ordinary ChPT, the systematic expansion
is effectively an expansion in the ratio
\begin{equation}
\label{mpofp}
\frac{M_\p^2}{(4\p F_\p)^2} \ .
\end{equation}
Since $M_\p^2$ is proportional to the quark mass $m$ at tree level,
the same is true for this ratio.
It follows that by taking $m$ small enough,
the ratio~(\ref{mpofp}) can be made arbitrarily small.
Since this is the ratio that controls the low-energy expansion,
this expansion is indeed systematic.

Similarly, under the power counting~(\ref{pcG}),
the ratio that should control the low-energy expansion
in the dilaton sector is (compare Eq.~(\ref{Md}))
\begin{equation}
\label{mdov}
\frac{M_\t^2}{(4\p v)^2} = \frac{b_0 (4-\D)}{4\p^2} \ .
\end{equation}
We illustrate this fact via several examples of dilaton self-energy diagrams
at next-to-leading order (NLO) and next-to-next-to-leading order (NNLO):
\begin{itemize}
\item {\em NLO tadpole}. This diagram comes from a single quartic vertex,
which provides an explicit factor of $M_\t^2/v^2$.
An additional factor of $(M_\t/4\p)^2$ comes from the loop integral.
\item {\em NLO one-loop diagram with two cubic vertices.}
Together, the two vertices provide a factor of $M_\t^4/v^2$,
while a factor $1/(4\p)^2$
comes from the loop, as the loop integral is dimensionless.
\item {\em NNLO double tadpole}. This diagram comes from a single $\f^6$ vertex.
The explicit factor is $M_\t^2/v^4$, and two additional factors of $(M_\t/4\p)^2$
come from the two tadpole loops.
\end{itemize}

The crucial question is thus the following.
Given the power counting~(\ref{pcG}),
is the ratio~(\ref{mdov}) parametrically small,
as is the ratio~(\ref{mpofp}) in ordinary ChPT,
or the ratio~(\ref{tauratio}) in dChPT?  The answer is No!
The reason why the ratio~(\ref{mdov}) fails to be parametrically small
is that it involves the parameter $b_0$.
As explained above, $b_0$ is the coupling of the $\c^4$ term
in the dilaton potential~(\ref{VbnG}),
and it is $O(1)$ because a pure $\c^4$ potential is scale invariant.
If we reexamine the expression for the dilaton mass~(\ref{Md}),
we see that the only reason why it is parametrically small
is the presence of the factor $v^2$, since $v$ itself is
parametrically small.  However, the factor of $v^2$ drops out
in the ratio~(\ref{mdov}).  Thus, would-be subleading EFT contributions,
such as the examples listed above, are all parametrically of the same size
as the leading order terms.

This negative conclusion implies that, in fact,
there is no systematic expansion for the dilaton sector of the
low-energy effective theory if the power counting~(\ref{pcG}) is assumed,
thereby ruling out this power counting for the pion-dilaton EFT.

There is, however, one exception to this conclusion.
The ratio~(\ref{mdov}) will become parametrically small,
if $\D$ is close to 4.  However, for this to be the case,
we need to assume that $\D-4$ is parametrically small,
in other words, we must replace the power counting~(\ref{pcG})
by the power counting~(\ref{pcD}), to which we turn next.

\subsection{\label{Dto4} $\D\to 4$ power counting}
The power counting~(\ref{pcD}) is based on the assumption that $\Osill$
is an irrelevant operator inside the conformal window, which becomes relevant
below the conformal sill, with $\D=4$ precisely at the sill.
We will thus assume that, below the sill,\footnote{%
  Both here and in Eq.~(\ref{Tnf}), which provides the basis
  for the power counting~(\ref{pcd}) of dChPT, one can replace $n_f^* - n_f$
  by $|n_f^* - n_f|^\eta$ for some $\eta>0$ without changing the basic
  physical features.  The actual value of $\eta$ is not known,
  and to avoid cumbersome notation we set $\eta=1$.
  \label{etaft}
}
\begin{equation}
\label{4mDelta}
4-\D \sim n_f^* - n_f \ .
\end{equation}

The $\cs$ spurion we have introduced for the $G\to 0$ power counting
is not appropriate for the $\D\to 4$ power counting, and we trade it
with the $\s$ spurion of Sec.~\ref{dchpt} according to
\begin{equation}
\label{Ssig}
\cs(x) = G e^{(4-\D)\s(x)} \ .
\end{equation}
Using the $\s$ transformation rule~(\ref{scales}) it is easy to check
that the transformation rule of the $\cs$ spurion is reproduced,
and thus that the action defined by the lagrangian~(\ref{totLS})
remains formally invariant, as it should.
As for the dilaton effective field of the EFT, it is convenient to similarly
switch to the $\t$ field, via
\begin{equation}
\label{chitotau}
\c(x) = f_\t e^{\t(x)} \ .
\end{equation}
Once again, using Eq.~(\ref{scalet}) it immediately follows
that the correct transformation rule of the $\c$ field is reproduced.

The reason why the $\s$ spurion is now the appropriate one
is that, according to Eq.~(\ref{Ssig}), this
spurion comes multiplied by the small parameter $4-\D$.
It follows that the $\D\to 4$ power counting works in the same way
as the power counting of dChPT.  As explained in Sec.~\ref{dchpt},
in dChPT the differentiation with respect to the $\s$ spurion
of the renormalized lagrangian~(\ref{Sren}) of the underlying theory
generates an insertion of the trace anomaly,
which in turn is parametrically small according to Eq.~(\ref{Tnf}).
Similarly, with the substitution~(\ref{Ssig}), here
the differentiation with respect to the $\s$ spurion
of the phenomenological lagrangian~(\ref{totLS})
generates one power of the small parameter controlling
the dilaton sector, which is now $4-\D$.
We recall that the large-$N_c$ extension of ChPT
that accommodates the anomalous U(1)$_A$ symmetry works similarly:
the $\th$ spurion plays the role of the $\s$ spurion,
and the differentiation with respect to $\th$ generates one power
of the small expansion parameter $1/N_c$  (see Sec.~\ref{dchpt}).

In fact, as we will now show, the $\D\to 4$ power counting is completely
equivalent to the power counting of dChPT, and the EFT it generates is just
dChPT itself.
Starting from the dilaton potential~(\ref{Vbn}),
which was valid for the $G\to 0$ power counting,
we expand out the spurion field $\cs$ as a power series in $\s$,
and the effective field $\c$ as a power series in $\t$.
The outcome is that the dilaton potential gets rearranged as
\begin{equation}
\label{V4mD}
V = f_\t^2 B_\t e^{4\t} \sum_{n=0}^\infty c_n (\t-\s)^n \ ,
\end{equation}
where the expansion coefficients satisfy
\begin{equation}
\label{cn}
c_n \propto (4-\D)^n = O(|n_f^* - n_f|^n) \ .
\end{equation}
Once again it is easy to check that the (formal) scaling dimension
of the potential remains 4.  As usual, the explicit breaking of
scale invariance is recovered by setting $\s=0$.

The similarity of Eq.~(\ref{V4mD}) to Eq.~(\ref{Vnf}) is evident:
In both cases the coefficient $c_n$ of $(\t-\s)^n$ behaves like
the small expansion parameter raised to the $n$-th power.
This implies that the $\D\to 4$ power counting reproduces
the dilaton potential of dChPT.  The same reasoning can be applied
to every term in the effective lagrangian, and the conclusion is
that the resulting EFT of the $\D\to 4$ power counting is indeed dChPT itself.

\subsection{\label{smr} Summary}
In this section we considered the two power countings that can be
motivated by the phenomenological lagrangian~(\ref{totL}).
For the $G\to 0$ power counting we found that both the VEV
of the $\c$ field and the dilaton mass $M_\t$ are proportional
to some (fractional) power of the small parameter $\tb_1$.
However, the parametric smallness of $M_\t^2$ originates solely
from its dependence on $v^2$.  Thus, in the ratio~(\ref{mdov})
that should control the low-energy expansion in the dilaton sector,
the dependence on the small parameter $\tb_1$ drops out.
The ratio comes out to be $O(1)$ as a direct consequence of the fact that
the coupling $b_0$ of the scale invariant dilaton potential is $O(1)$,
and this invalidates the $G\to 0$ power counting.

As for the $\D\to 4$ power counting, we showed that this is the same
power counting as in dChPT.  However, the power counting of dChPT
is better motivated because it is derived directly from the underlying theory,
rather than from a phenomenological lagrangian.  The VEV of the $\t$ field
in dChPT behaves in a qualitatively different way.  The $O(1)$ coupling
of the scale invariant dilaton potential, $c_0$, leads to a VEV that is
inversely proportional to the small parameter $n_f^*-n_f$ (see Eq.~(\ref{vd})).
For this reason, one needs to make (mild) assumptions
about the global form of the potential, as we have discussed in detail
in Ref.~\cite{LMP}.  Once these assumptions are satisfied one can
use the $\t$ shift to eliminate the $O(1)$ part of $c_0$,
a feature which implies that $c_0$ is not physical in dChPT.
The resulting low-energy expansion is systematic.

Two EFTs that are so qualitatively different cannot both describe
the same underlying theory.  We have been able to rule out the EFT of
the $G\to 0$ power counting, because it does not generate a systematic
low-energy expansion.  We conclude that the correct EFT with the
correct power counting is dChPT.

\subsection{\label{Dpot} $\D$-potential}
The so-called $\D$-potential was used in a numbers of publications
as the leading-order dilaton potential
in fits to numerical data \cite{LSD3fit,AIP1,AIP2,AIP3,AIPuniverse,AIPLSD}.
It was argued in Appendix~A of Ref.~\cite{AIP3} that the $\D$-potential
arises from the $G\to 0$ power counting.\footnote{%
The role of the lagrangian~(\ref{totL}) as the underlying lagrangian
responsible for the power counting of the EFT
was most explicitly described in the last paragraph of Ref.~\cite{AIPLSD}.
}
The arguments in Ref.~\cite{AIP3} are rather opaque to us,
as we have pointed out in Ref.~\cite{KMIfit}.  Nonetheless, as it turns out
it is easy to check the claim that the $\D$-potential can be
used at leading order in a systematic expansion.  In our notation,
the $\D$-potential reads
\begin{equation}
\label{VDelta}
V_\D = \frac{M_\t^2}{4(4-\D) \vevc^2}
\left( \c^4  - \frac{4}{\D} \frac{\c^\D}{\vevc^{\D-4}} \right) \ .
\end{equation}
Upon substituting Eq.~(\ref{Md}) for the dilaton mass,
and Eq.~(\ref{vsolve}) for the VEV of the $\c$ field,
the leading-order dilaton potential of the $G\to 0$ power counting,
Eq.~(\ref{Vdn}), is readily recovered.

As we showed above,
the $G\to 0$ power counting does not lead to a systematic low-energy expansion,
and thus this is true for the $\D$-potential as well.
The only way out is to take $4-\D$ small, but this amounts to reverting
to the $4-\D$ power counting, at which point the pion-dilaton EFT
becomes dChPT.

We comment that one might go ahead and postulate that for some
unknown reason the dilaton mass is small compared to the VEV of the $\c$ field.
As we have shown above (note in particular Eq.~(\ref{mdov})),
if $4-\D$ is not small
this is equivalent to the {\it ad hoc} assumption that the
coupling $b_0$ of the scale-invariant quartic potential in the EFT
is small; whereas physically, there is nothing in the underlying theory
to prevent $b_0$ from being $O(1)$.\footnote{%
This {\it ad hoc} fine tuning was most explicitly invoked
in the last paragraph of Sec.~3.1 of Ref.~\cite{AIPuniverse}.}

While we cannot rule out that $b_0$ could be {\em accidentally} small,
this is {\em not} the same as having a systematic low-energy expansion.
The premise of starting from Eq.~(\ref{totL}) is that scale invariance is restored
at the conformal sill because $G\to 0$ (if the parameter $\D$ is to be kept as
a free parameter).   Accordingly, the dilaton mass would be expected to
vanish at the sill. As we showed here,
while the dilaton mass indeed tends to zero, the ratio~(\ref{mdov}) remains
constant and $O(1)$ when $G\to 0$,
invalidating the $G\to 0$ power counting
because this ratio controls the expansion in the dilaton sector.
One can make an {\it ad hoc} attempt to ``rescue'' the EFT by taking $b_0$
to be accidentally small, but this does not follow from Eq.~(\ref{totL}),
and, moreover, this does not remedy the problem that the ratio~(\ref{mdov})
would not tend to zero when the conformal sill is approached.\footnote{%
For the situation in dChPT, see discussion around Eqs.~(\ref{tauratio})
and~(\ref{ratiohscale}).}

The assumption that $b_0$ is small would thus turn the $\D$-potential
into a model; the smallness of the mass of the dilaton
(and of the ratio~(\ref{mdov})) is no longer
explained from the dilaton being a pseudo-NGB. Thus, the claimed connection
of the EFT with the underlying gauge theory, and even with the
phenomenological lagrangian~(\ref{totL}) where the explicit breaking
of scale invariance comes from the term $G\Osill$, is also lost.
While the $\D$-potential
model might be useful, it does not qualify as an effective field theory.

Returning to dChPT, the coefficient $c_0$ of the quartic potential
is assumed to be $O(1)$, as $b_0$ here. This simply follows from the
quartic potential being scale invariant.
But as we explained in detail in Sec.~\ref{dchpt},
in dChPT the coefficient $c_0$ is unphysical
as the $\tau$-shift symmetry can be used to set $c_0$
to any desired value.  For this to work we need some
mild assumptions about global features of the dilaton potential \cite{LMP}
that are consistent with the power counting underlying dChPT, and do not
contradict it---unlike the assumption that $b_0$ is unnaturally small,
which is entirely unrelated to any power counting originating from
the underlying theory and/or from the
phenomenological lagrangian~(\ref{totL}).

\section{\label{disc} Discussion}
An effective low-energy field theory must reproduce the correlation functions
of the underlying theory, and it must do so systematically,
order by order in the expansion in its small parameters.
These stringent requirements explain why the construction of an EFT
must follow very strict rules, as we briefly reviewed in Sec.~\ref{dchpt}.

The underlying theories we have been dealing with in this paper are QCD-like.
Such theories are well understood theoretically.  What we mean by this is that
they are defined by a renormalizable lagrangian, and, thanks to their
asymptotic freedom, they can be defined non-perturbatively on the lattice
by approaching the continuum limit at the gaussian fixed point.
The safe approach to the construction of a pion-dilaton EFT
is thus to match the EFT directly to the underlying theory, as we have done.
(As in the derivation of ordinary chiral perturbation theory
from lattice QCD, a key intermediate step is the identification of the
underlying continuum lagrangian with the Symanzik effective theory
at leading order in an expansion in the lattice spacing.)
The approximate scale symmetry stems from
the smallness of the beta function in ``walking'' theories
which are close to the conformal window.  The distance in theory space
to the sill of the conformal window thus naturally serves to provide
the small parameter which controls the mass and the
interactions of the dilatonic meson.

The construction of a pion-dilaton EFT is significantly more
complicated than ordinary ChPT.  As it turns out, a key difficulty
is that a purely quartic potential for the (effective) dilaton field
is allowed by scale invariance.  Hence, the power counting of
the pion-dilaton EFT must allow the quartic potential to be present
with an $O(1)$ coupling in the low-energy lagrangian.
The question is thus whether the EFT can cope with the quartic potential
and still generate a systematic low-energy expansion.
As we have shown in Refs.~\cite{GSEFT,LMP}, in dChPT the answer is positive.
At the technical level, the freedom to shift the dilatonic meson field $\t$
plays an important role in ensuring that the $O(1)$ coupling
of the quartic potential does not jeopardize the systematic expansion.
As an example, we reviewed how the ratio controlling
the low-energy expansion in the dilaton sector, Eq.~(\ref{tauratio}),
is parametrically small, as it should be.

Our main goal in this paper was to explore power countings
for the pion-dilaton EFT which are motivated by the phenomenological picture
of the conformal transition presented in Sec.~\ref{phenosill}.
We note that, from the start, there are serious objections to this program.
First, we have already shown that the pion-dilaton EFT can be constructed
by matching it directly to the underlying renormalizable theory.
Why should we, then, trade the actual underlying theory with
a speculative phenomenological lagrangian
which may or may not reflect the physics of the underlying theory?

The second objection is that this phenomenological approach
is in fact ambiguous, in that it suggests {\em two} alternative
power countings for the pion-dilaton EFT.  In one sentence, one assumes that
the phenomenological lagrangian consists of a conformal part,
plus a single term $G\Osill$.
The conformal transition is triggered
either by the coupling $G$ of $\Osill$ tending to zero,
which leads to the first power counting; or, alternatively,
by the scaling dimension $\D$ of $\Osill$ tending to 4,
which leads to the alternative power counting, with a potentially different EFT.
However, since the underlying theory is well defined, there cannot exist
two different EFTs describing the low-energy physics.  Any assumption that
leads to two different possible EFT constructions thus leads us to an apparent
paradox:  which EFT is the correct one?  Only one EFT can match the
correlation functions of the underlying theory.

To gain more insight,
in Sec.~\ref{phenosill} we went ahead and investigated the resulting EFTs
that follow from these two power countings,
setting aside the questions raised above.
We found that the $G\to 0$ power counting fails to generate
a systematic expansion.  This failure is directly related
to the presence of the quartic dilaton potential with the unsuppressed
$O(1)$ coupling.  Unlike in dChPT there is no mechanism to ``tame''
this $O(1)$ coupling.  As a result, the ratio~(\ref{mdov}) which defines
the would-be expansion parameter of the dilaton sector turns out to be $O(1)$,
instead of being parametrically small as required.
This implies an inconsistency of the EFT defined by the $G\to 0$ power counting.
Furthermore, since, as we have shown the $\D$-potential is completely equivalent
to the leading-order potential of the $G\to 0$ power counting,
this conclusion applies to the $\D$-potential as well.

As for the $\D\to 4$ power counting,
we found that it is in fact the same as the power counting of dChPT,
and the resulting EFT is just dChPT itself.

Thus, setting aside more fundamental possible objections to the
phenomenological lagrangian~(\ref{totL}), we see that
it does not lead to any new option.  This is as one might have expected.
Truly different power countings would lead to different EFTs,
and thus to different physical predictions,
leading to the paradox described above.
Our results resolve this paradox, because what we find is that there is only
a single consistent power counting for the pion-dilaton EFT, even if we
start from the phenomenological lagrangian~(\ref{totL}),
and the resulting EFT is dChPT.

\vspace{3ex}
\noindent {\bf Acknowledgments}
\vspace{2ex}

We thank the referee for their insightful questions that have led
to improving the clarity of this paper.
This material is based upon work supported by the U.S. Department of
Energy, Office of Science, Office of High Energy Physics,
Office of Basic Energy Sciences Energy Frontier Research Centers program
under Award Number DE-SC0013682 (MG).
YS is supported by the Israel Science Foundation under grant no.~1429/21.

\vspace{5ex}
\def\refttl#1{\medskip\noindent \underline{#1}}
\def\refttl#1{}

\end{document}